\documentclass[reprint,superscriptaddress,twocolumn,aps,prb,showpacs,floatfix]{revtex4-1}
\usepackage{graphicx}%
\usepackage{dcolumn}%
\usepackage{bm}
\date{\today}
\usepackage{hyperref}
\begin{document}

%=============================================INTRODUCTION===============================================
 
\title{Fabrication of superconducting tantalum nitride thin films using infra-red pulsed laser deposition}
\author{S. Chaudhuri} \email{sachaudh@jyu.fi}\affiliation{Nanoscience Center, Department of Physics, University of Jyv\"askyl\"a, P. O. Box 35, FIN-40014 University of
Jyv\"askyl\"a, Finland} 
\author{L. Chandernagor}\affiliation{Nanoscience Center, Department of Physics, University of Jyv\"askyl\"a, P. O. Box 35, FIN-40014 University of
Jyv\"askyl\"a, Finland} \affiliation{ENSICAEN,  6 bd Mar\'echal Juin, F-14050 CAEN Cedex 4 France}
\author{M. Lahtinen }\affiliation{ Department of Chemistry, Laboratory of Inorganic and Analytical Chemistry, P.O. Box 35, FIN-40014 University of Jyv\"askyl\"a, Finland}
\author{M. Ging}\affiliation{Nanoscience Center, Department of Physics, University of Jyv\"askyl\"a, P. O. Box 35, FIN-40014 University of
Jyv\"askyl\"a, Finland} \affiliation{ENSICAEN,  6 bd Mar\'echal Juin, F-14050 CAEN Cedex 4 France}  
\author{I. J. Maasilta}  \affiliation{Nanoscience Center, Department of Physics, University of Jyv\"askyl\"a, P. O. Box 35, FIN-40014 University of
Jyv\"askyl\"a, Finland}
 
%=============================================Abstract===============================================

\begin{abstract}
We report the successful fabrication of    superconducting tantalum nitride (TaN) thin films using a  pulsed laser deposition technique with 1064 nm radiation. Films with  thickness  $ \sim $ 100 nm deposited on MgO (100) single crystals and on oxidized silicon  (SiO$ _{2} $) substrates  exhibited a superconducting transition temperature of  $\sim  $ 8 K and 6 K, respectively. The topography of these films were investigated using atomic force and scanning electron microscopy, revealing fairly large area particulate free and smooth surfaces, while the structure of the films were investigated using standard $ \theta $-2$ \theta $ and glancing angle   X-ray diffraction techniques. For films grown on  MgO a face-centered cubic phase of TaN was observed,  while films grown on SiO$ _{2} $ exhibited the face-centered cubic and as well as a mononitride hexagonal phase. The transition temperature  of the TaN deposited on SiO$ _{2} $ was found to be more sensitive to the  nitrogen pressure  during deposition as compared to the TaN deposited on MgO.
\end{abstract}
\maketitle
%=============================================INTRODUCTION===============================================

\section{Introduction}
Efficient fabrication of materials with intermediate and high  superconducting transition temperature ($ T_C $) is essential for the  development and wide-spread use  of  superconducting electronics. Particularly attractive are those materials whose $ T_C $ is above 4.2 K, the boiling point of liquid He$ ^{4} $.  There exists a host of materials fulfilling this criteria ranging from elementary metals like Nb ($ T_C $ $ \sim $ 9 K) to more complex multi-element based copper oxide materials ($ T_C $ up to $ \sim $ 135 K) \cite{Poole}. However, the superconducting properties of such multi-element  materials are  extremely   sensitive to stoichiometry, and thus their fabrication is non-trivial. A good compromise in terms of $ T_C $ and ease of fabrication are   metal nitrides as they are binary in composition but their $ T_C $s  are higher  than their elemental counterparts. Particularly, the nitride of  tantalum (TaN) in thin film form  is a promising material for  micro-electronic device applications.  In its normal state, TaN has earned the reputation as an excellent diffusion  barrier  material\cite{wittmer:540, holloway:5433}, as resistors with low temperature coefficient\cite{katz:5208}  and as hard coatings\cite{Baba1996429},  while  superconducting TaN was recently demonstrated to be  a much better candidate   than NbN for  single photon detection \cite{SSPD} because of its lower gap and lower density of states. TaN also works well as a normal state barrier in SNS Josephson junctions \cite{SNSAPL,minna}.  

Depending on the amount of incorporated nitrogen, $ x $, the tantalum nitride system TaN$ _x $  can exhibit a variety of crystallographic phases like cubic, hexagonal  or tetragonal\cite{JJAP.10.248,APA}. Among these, the  hexagonal and tetragonal  phases are thermodynamically stable\cite{PhysRevB.71.024111}, but with  no signature of superconductivity down to 1.5 K\cite{reichelt:5284, PhysRev.93.1004}. The  stoichiometric mononitride  (TaN) phase  possess a face-centered cubic (FCC) structure and exhibits superconductivity with a $ T_C $  of 8.15 K\cite{kilbane:107}, which  under special preparation conditions, has been pushed up to 10.8  K\cite{reichelt:5284}. Thermodynamically this phase is  stable only at high temperatures $ \sim $  1800 $^{\circ}$C or under high pressures \cite{Boiko},  but it is metastable at room temperature. Nevertheless, the metastable FCC phase can be obtained at low temperatures  for TaN thin films prepared by plasma assisted or ion irradiated process, whereby the energetic ions  provides the necessary energy of crystallization for the FCC phase formation\cite{ensinger:6630}. TaN$ _{x} $ thin films  have been fabricated mainly  by  reactive sputtering  \cite{Gerstenberg01011964, holloway:5433,olowolafe:4099,Shin,APA,rossnagel:2328,Ilin,kilbane:107,SSPD} and nitrogen implantation of evaporated tantalum\cite{ensinger:6630,JJAP.10.248,Baba1996429,Arranz}  on a variety of substrates. In most cases the fabricated films were intended for normal state applications and therefore the superconducting aspects were not investigated. So far, superconducting TaN thin films have been prepared on sapphire\cite{SSPD}, glass\cite{kilbane:107}, quartz\cite{Gerstenberg01011964}  and alumina \cite{kilbane:107} substrates, all using reactive sputtering.  TaN thin films (with no reported superconductivity)   have also been fabricated using pulsed laser deposition (PLD) technique where  the first harmonic  (532 nm radiation)  from a Nd:YAG laser was used as the source  \cite{HK, matenoglou}. The first harmonic has been traditionally used with Nd:YAG lasers in PLD, however it has the disadvantage that the output power is much reduced compared to the fundamental 1064 nm radiation. We have recently shown that it is actually possible to obtain high quality superconducting NbN films  using just the  fundamental 1064 nm  radiation  from a Nd:YAG laser\cite{nbn}, and the same tactic was used in this work for TaN deposition.   
 
In this paper, we report the successful fabrication of superconducting TaN thin films on (100) oriented single crystals of MgO and oxidized (100) silicon (SiO$_2 $) substrates. The films were fabricated by the ablation of a high purity Ta  target in a ultra high purity nitrogen environment   using the  fundamental  1064 nm laser pulse emission from a Nd:YAG laser. We have tried to maximize the $ T_C $  by carefully   optimizing the  growth parameters.  Since our long term goal is the fabrication of superconducting TaN based tunnel junction devices, in the  present study we have  investigated the topographical, structural and the electrical aspects of our thin films.
Our recent results on NbN\cite{NbN2} and Nb\cite{Nb} based tunnel junction micro-devices hint that  intermediate $ T_C $ based thin film superconductors, such as TaN,  may be the key to  the practical realization of high operation temperature (1 K) electronic coolers\cite{RMP,muhonen}.

%=============================================Experimental=============================================

\section{Experimental}
TaN thin film of thickness $  \sim$ 100 nm were fabricated by ablating a high purity (99.9\%) Ta target in an ultra high purity nitrogen  environment. For ablation, 4 ns laser pulses corresponding to   the fundamental ($ \lambda $ =1064 nm, $ h\nu $= 1.16 eV) emission of  an Nd:YAG laser were used.  The details of the 1064 nm radiation based PLD technique  can be found elsewhere\cite{nbn}. The laser was operated at a frequency of 10 Hz, the substrate to target distance was 5 cm while the incident energy density was $ \sim $ 6 Jcm$ ^{-2} $.  Typical growth rate was $ \sim $ 4 nm/s. TaN films were deposited on two types of substrates  : (100) oriented single crystals of MgO (purchased from Crystec GmbH) and  thermally oxidized (100) silicon substrates (oxide thickness $ \sim $300 nm). Since for any potential device application the leakage current to the substrate has to be minimized, oxidized silicon was chosen over bare silicon due to the excellent dielectric properties of silicon oxide thin films. The deposition temperature ($ T_D $) was kept fixed at  700 $^{\circ}$C, as previous studies suggested that higher $ T_D $ favours the formation of larger grains \cite{HK}. The nitrogen pressure ($ p_N $) was varied to change the Ta:N ratio in the films.  After deposition, the films were let to cool back to room temperature in a constant flow of high purity nitrogen at the same pressure as at which the TaN was deposited.  Also, films corresponding to the same pressure but different substrates  were fabricated during the same deposition process. Room temperature atomic force microscopy (AFM)   was carried out in the tapping mode using  a Nanoscope V AFM. The X-ray diffraction analyses on the films were made with PANalytical X'Pert Pro alpha 1 diffractometer using glancing angle of 5$^{\circ}$ and Johansson monochromatized Cu K$_{\alpha 1}$   radiation ($\lambda$ = 1.5406  \AA ; tube settings: 45 kV, 30 mA). For each measurement, the sample plate was prepared on a  zero-background holder made from silicon. The data was collected from a spinning sample by X'Celerator detector in the 2$  \theta$-range of 10-110$^{\circ}$ with a step size of 0.017$ ^{\circ}$ and counting times of 1000 s (480 s for glancing angle) per step. The diffraction data was handled by X'Pert HighScore Plus v. 2.2d software package and the ICDD PDF 4+ powder diffraction database was used for the qualitative search-match phase identification. Resistivity measurements were carried out in a four probe configuration using a dipstick immersed in a liquid helium bath, with a current bias of 10 $  \mu$A. 
 
 %=============================================Results===============================================
\section{Results}
\begin{figure}[hbtp]
\includegraphics[width=1\columnwidth]{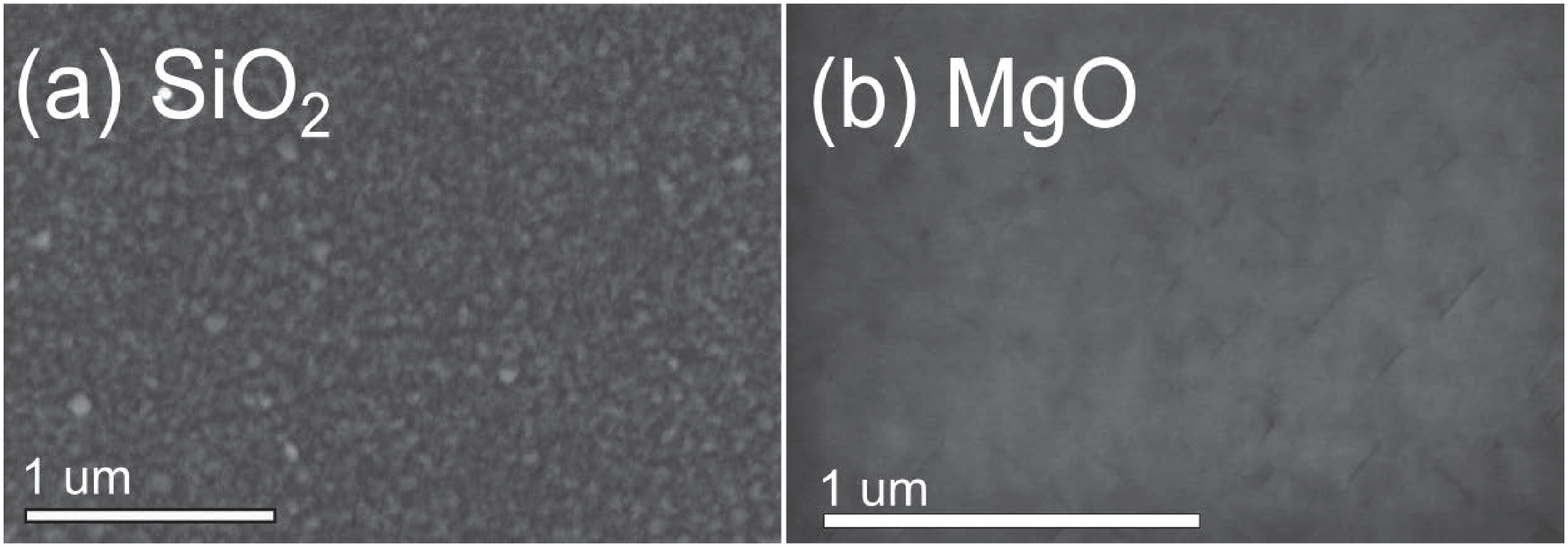}
\caption{Room temperature scanning electron micrograph of TaN grown on (a) SiO$_2$ and (b) MgO respectively. The length scale is indicated by the horizontal white line on bottom left.}\label{Fig1}
\end{figure}
Scanning electron microscopy images on TaN grown on both SiO$ _2 $ and MgO  reveal a very low particulate density, as seen from Fig. ~\ref{Fig1}. It is easy to obtain particulate free regions that are up to several tens of microns in dimension. The grainy texture of the TaN  film on SiO$ _2 $ is clearly visible, while the film on MgO appears much more uniform.

\begin{figure} 
\includegraphics[width=0.7\columnwidth]{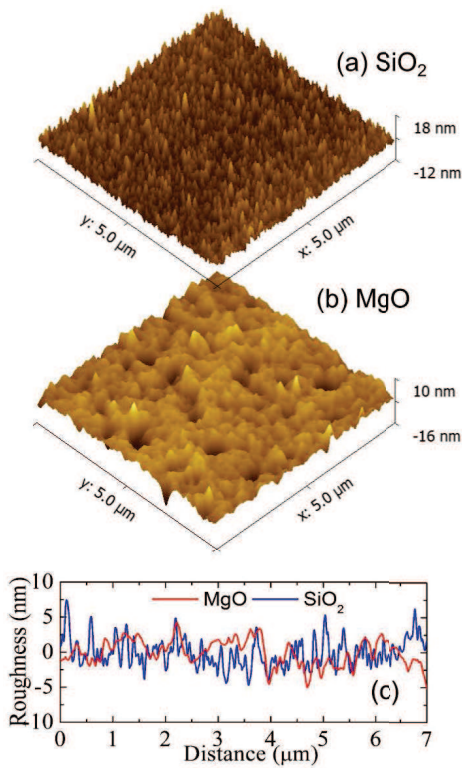}
 \caption{[Color online] Room temeperature atomic force microscopy images of TaN thin film grown on  (a) SiO$ _2 $ and   (b) MgO  substrates deposited at the same time at  $ p_N $ = 55.5 mTorr. (c) Line scans repersenting the variation in surface roughness measured along one of the diagonals of the scanned area.  }\label{Fig2}
\end{figure} 

The surface topography of the TaN films were investigated using atomic force microscopy, with the results for 5$\times$5 $ \mu $m$ ^{2} $ scans shown in Fig.~\ref{Fig2}. The root mean square  surface roughness for the films grown on  MgO and  SiO$ _2 $  were 2.2 and 2 nm, respectively, with  the  line scans showing the roughness of the surface measured along one of the diagonals of the scan area are shown in Fig.~\ref{Fig2}(c).    It is clearly seen that the morphologies of the films grown on the two different substrates are quite different from each other. The spatial scale of the  surface roughness of the film grown on   SiO$ _{2} $ is much smaller than the corresponding one in the  film grown on MgO. These TaN films tend to be smoother than NbN films of similar thickness prepared in the same PLD chamber, which  revealed  an rms surface roughness of  $ \sim $ 3 nm \cite{nbn}. Using the rougher NbN films, we have already  successfully fabricated NbN based tunnel junction micro-thermometers\cite{NbN2} capable of operating from 0.1 K upto  $ T_C $ of NbN\cite{NbN2}. Thus, the low particulate density in conjunction with the intrinsic smoothness of the surface   is very promising for the fabrication of TaN based micro scale  tunnel junction devices.  

\begin{figure} 
\includegraphics[width=1\columnwidth]{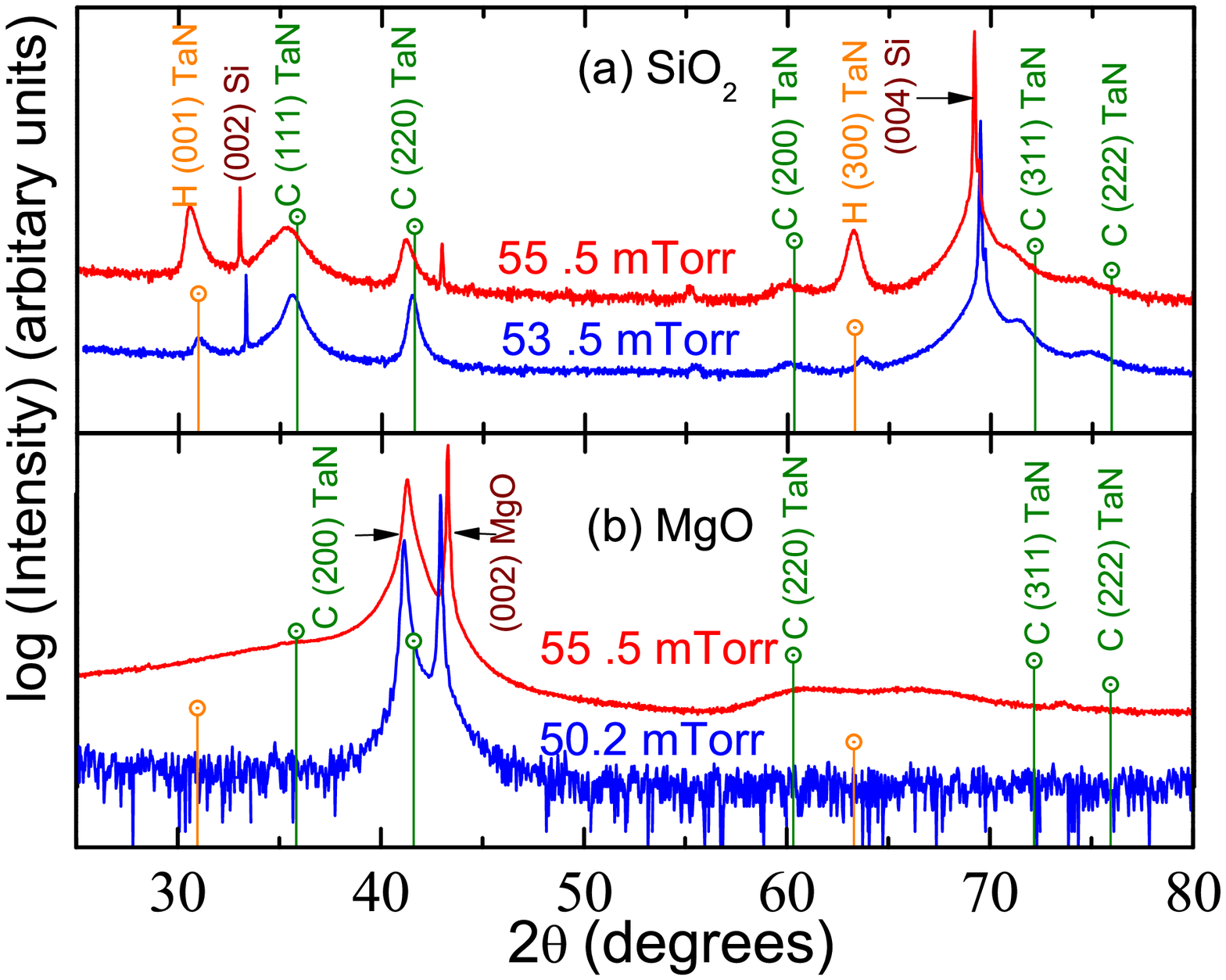}
\includegraphics[width=1\columnwidth]{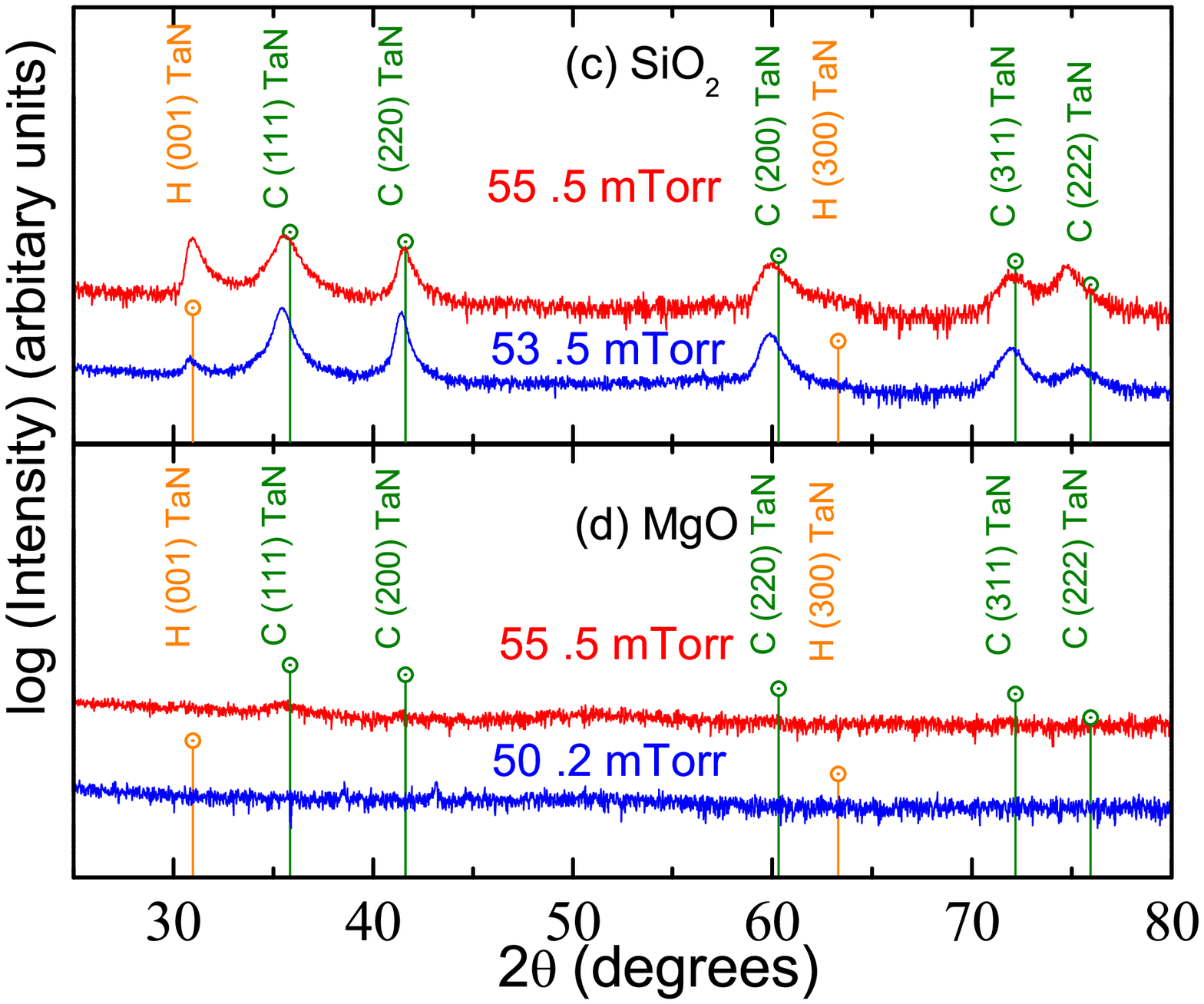}
\caption{[Color online] Room temperature X-ray diffraction patterns of TaN films deposited on (a) oxidized (100) Si and (b) (100) MgO substrates  measured in  $ \theta $-  2$ \theta $ mode. In each case diffractograms corresponding to $ p_N $ = 55.5 (red) and 53.5 (blue)   mTorr on SiO$ _2 $  and 55.5 (red) and 50.2 (blue) mTorr  on MgO is shown.  The corresponding glancing angle measurements are shown in panels (c) and (d). The reflections from various lattice  planes of cubic ($ C $) and hexagonal ($ H $) phases have been identified  and marked, with the characteristic positions of the cubic and hexagonal  TaN phases indicated by the  green and orange tick marks, respectively.}
\label{Fig3}
\end{figure} 

Structural analysis of the TaN thin films, carried out  at room temperature using the traditional $ \theta-2\theta $ scanning method,  revealed diffraction peaks originating both from the substrate and  the TaN film. As seen from Fig.~\ref{Fig3} (a) and (b), the  diffractograms are dominated by the respective substrate peaks. Results of glancing angle measurements, where only the peaks from the films are observed, are displayed in Fig.~\ref{Fig3} (c) and (d).  Analyzing both the  $ \theta-2\theta $ and the glancing angle data, we find that TaN films  grown on the SiO$ _2 $ substrate exhibit  distinct peaks corresponding to (111), (200), (220), (311) and (222) planes of  a face centered cubic (FCC) structure.   The simultaneous occurrence of reflections from all these planes indicates that our TaN film on SiO$ _2 $ consists of FCC TaN grains with random orientations.   TaN thin films grown  on silicon and glass substrates at low temperatures have also been shown to exhibit such a poly-crystalline nature \cite{APA,ensinger:6630}. However, the films grown on the MgO substrate  exhibit a dominant (200) reflection indicating that there is a preferred growth direction. It has been observed that when TaN was deposited using metal evaporation under simultaneous  nitrogen ion irradiation,  the preferred growth direction changes from  (111) to (100) with increasing irradiation intensity  \cite{ensinger:6630}. 

Using the Bragg formula, $ 2d\sin\theta =n\lambda $, we obtain a value for the out-of-plane  lattice parameter, $ c  \sim $ 4.4~\AA ~for the  FCC phase of the TaN films grown on both  MgO and SiO$ _{2} $.  In addition to the peaks associated with the FCC phase, the diffractograms  of TaN grown on  SiO$ _{2} $  exhibit other peaks that possibly correspond to reflections from (001) and (300) planes of the hexagonal TaN phase, which has also been observed for TaN grown on (111) \cite{Li} and (100) \cite{HK} Si substrates  at a high temperature and substrate bias, respectively.   In our case, the peak occurring at $ 2\theta $ $ \sim $ 30.6$^{\circ}$ corresponds to an out-of-plane lattice parameter $ c $ = 2.92~ \AA, close to what is expected for hexagonal $ \epsilon $-TaN ($ c$ = 2.91 \AA) \cite{JJAP.10.248}. Similar to FCC TaN, this hexagonal phase is  also  stoichiometrically mononitride, thus the identical stoichiometry is  expected to lead to the co-existence of the two phases.  The peak occurring at $ 2\theta $ $ \sim $ 63.2$^{\circ}$ corresponds to an in-plane lattice parameter $ a  \sim $ 5.07 \AA,   close to the expected value of   5.19 \AA  for the  hexagonal $ \epsilon $-TaN \cite{JJAP.10.248}.  Thus, just as the FCC phase, the hexagonal phase  is polycrystalline, as well. It can also be seen from Fig.~\ref{Fig3}(a) and (c) that the peak intensities of the hexagonal TaN  phase diminish with decreasing $ p_N $. 

 For both the FCC and hexagonal structures the TaN peaks are broad, indicating a small crystallite size $ w $. Using the Debey-Scherrer equation, $ w $ = $0.9 \lambda/( \beta  \cos\theta)$, where $ \beta $ is the FWHM of a single diffraction peak, we estimated the size of the crystallites, as listed in Table~\ref{tab}. For films grown on SiO$ _{2} $  at 53.5 mTorr and  on MgO  at 50.2 mTorr, the peaks corresponding to the FCC TaN phase are slightly narrower than the corresponding  films grown at 55.5 mTorr, indicating a larger crystallite size. For example on SiO$ 2 $,  the crystallite sizes vary from $ \sim $ 83 to 224~\AA~ for $ p_N $ = 53.5 mTorr,   in contrast to  $ \sim $ 71 to 165~\AA~  for $ p_N $ = 55.5 mTorr.  Similarly, the FCC crystallite size of TaN grown on MgO varies from 206 - 453 ~\AA~ for  $ p_N $ = 50.2 mTorr,  to 352 - 817~\AA~ for    $ p_N $ = 55.5 mTorr.    In all cases, the (200) oriented crystallites possess the largest size, ranging from  $\sim  $  453 - 817 ~\AA~ on MgO to 165 - 244 ~\AA~ on SiO$ _2 $. Interestingly, in most cases  $ w $ is larger than the  observed coherence length in some TaN thin films   $ \sim $  90 ~\AA~  \cite{PhysRevB.86.014514}. 

 In case of the MgO samples,  the extremely broad cubic TaN peaks (111) and (220) disappear as the pressure decreases from  55.5 to 50.2 mTorr, as seen in Fig. ~\ref{Fig3} (b).   Furthermore, the clear absence of the FCC peaks in the glancing angle data of TaN on MgO indicates that the very surface of the films grown on MgO is amorphous, and the crystalline phases  exists only closer to the substrate-film interface. These peaks start to be visibile only when the glancing angle is increased  to 8$ ^\circ $. Similarly, the  hexagonal (300) peak of TaN on   SiO$ _{2} $ is also absent in the surface region. The fact that the surface of the TaN film on MgO is amorphous, and that the  hexagonal  and the cubic phases differ in grain size and morphology \cite{APA} are  possibly the reasons behind  the different  observed morphologies of the films grown on  MgO and SiO$ _{2} $ in our AFM measurements. 
\begin{table}
\begin{tabular}{ccccccc}
\hline 
\hline
Peak position  & Miller & Crystal  & \multicolumn{4}{c}{Crystallite size (\AA)}\tabularnewline
(2$\theta$) &   indices& structure  & \multicolumn{2}{c}{SiO$ _2 $} (mTorr) & \multicolumn{2}{c}{MgO} (mTorr)\tabularnewline
(degrees) &  &  & 55.5 & 53.5 & 55.5 & 50.2\tabularnewline
 \hline
35.58 & 111 & Cubic & 71 & 121 &  & \tabularnewline
41.57 & 200 & Cubic & 165 & 224 & 453 & 817\tabularnewline
60.4 & 220 & Cubic & 84 & 83 &  & \tabularnewline
72.06 & 311 & Cubic & 90 & 90 &  & \tabularnewline
89.23 & 400 & Cubic & 146 & 135 & 206 & 352\tabularnewline
30.74 & 001 & Hexagonal & 185 & 157 &  & \tabularnewline
63.24 & 300 & Hexagonal & 196 & 181 &  & \tabularnewline
 &  &  &  &  &  & \tabularnewline
 \hline
 \hline
\end{tabular}\caption{Crystallite sizes associated with various orientations of TaN grains grown on SiO$_2 $ and on MgO substrates extracted from our X-ray data. }\label{tab}
\end{table}
\begin{figure} 
\includegraphics[width=1\columnwidth]{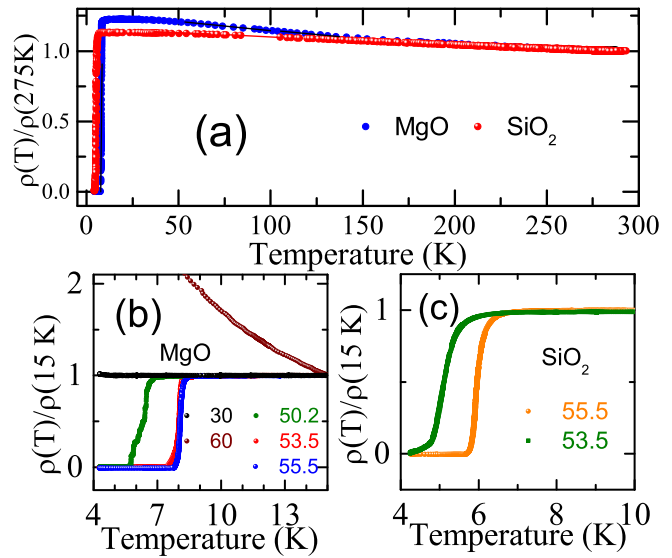}
\caption{[Color online] (a) Temperature dependence of the resistivity normalized to the values at 275 K, of TaN film grown  at 700$ ^{\circ} $C, revealing a superconducting transition of $ \sim $ 8 K and 5 K on  MgO and SiO$ _2 $ substrates, respectively. The respective  $ p_N $  were  55.5 and 53.5 mTorr. Temperature dependence of the resistivity of a TaN film,  normalized to its value at 15 K, grown   on (b)  MgO and   (c)  SiO$ _2 $ at several values of $ p_N $ indicated. Th highest obtained $ T_C $ were 8 K and 6 K on MgO and SiO$ _2 $, respectively.  }\label{Fig4}
\end{figure} 

The variation of  the normalized resistivity of two  TaN films  with good $T_C$ values are shown in Fig.~\ref{Fig4} (a), measured from room temperature down to liquid helium temperature 4.2 K.  The film on  MgO was grown at  $ p_N $ = 55.5 mTorr, whereas for the film on SiO$ _2 $,   $ p_N $ =  53.5 mTorr.   The room temperature resistivity value were $  \sim$ 50 and 150  $\mu\Omega $cm on the MgO and SiO$ _2 $,  respectively.  We see that upon cooling from  room temperature, both films exhibited a negative temperature coefficient behaviour all the way down till $ T_C $. The residual resistivity ratio  $ \rho(275K)/\rho(20K) $ was $ \sim $ 0.7 - 0.8.  In  Fig.~\ref{Fig4} (b) and (c)   we show the resistive behaviour of films grown on MgO and  SiO$ _2 $, respectively, at various $ p_N $ . For films grown on MgO, at $ p_N $  = 30 mTorr, the resistivity remained almost constant with no signature of $ T_C $ down to 4.2 K. Upon increasing $ p_N $ to  50.2 mTorr, a superconducting transition occurs at $ T_C $ $ \sim $ 6 K, which can be pushed up to $ \sim $ 8 K when $ p_N $ =  53.5 - 55.5 mTorr. At  $ p_N $ = 60 mTorr, an insulating phase is observed with  no signature of $ T_C $ down to 4.2 K.  On the SiO$ _2 $ substrate, superconductivity was observed  only  in films grown at $ p_N $ =  53.5 - 55.5 mTorr. So far, the highest values of $ T_C $ obtained on MgO and SiO$ _2 $ were  grown at   $ p_N $ = 55.5 mTorr, giving $T_C$ values of $ \sim $ 8 K and 6 K for MgO and SiO$ _2 $ substrates, respectively. The corresponding widths of the transitions were  about 0.25 and 0.4 K, respectively.  

Although the lattice constants of the FCC phases of TaN on MgO  and SiO$ _2 $ are identical,  their $ T_C $ values differ  significantly. We speculate that this is possibly because on  SiO$ _2 $, due to the exsistence of the two different phases (superconducting FCC and non-superconducting hexagonal), the effective thickness of the superconducting FCC phase is less than in the MgO counterpart. It is well know  that the $ T_C $ of thin films is thickness dependent, whereby  $ T_C $  typically decreases  with decreasing film thickness\cite{kang:033908}.  On the other hand, a comparison of the films grown on   SiO$ _2 $ at 53.5 and 55.5 mTorr (Fig.~\ref{Fig4}(c))  shows that although  the hexagonal phase fraction is comparatively lower and FCC grain size is larger in the film grown at 53.5 mTorr (Fig. \ref{Fig3}(c)), its  $ T_C $  is  about 1 K lower than the for the film grown at $ p_N $= 55.5 mTorr.  A similar situation also arises for films grown on MgO. In this case,  the film grown at $ p_N $  = 50.2 mTorr possess a  larger FCC grain size and a reduced $ T_C $     as compared to film grown at $ p_N $= 55.5 mTorr (Fig.~\ref{Fig4}(b)). Therefore, the grain size might play an important role in determining the  $ T_C $ of our films\cite{T1s,JPSJ.41.1234,PhysRevLett.17.632}.
 \section{Conclusions}
In conclusion, we have successfully fabricated superconducting tantalum nitride thin films on magnesium oxide and oxidized silicon substrates using a pulsed laser ablation technique with infrared wavelength of 1064 nm.  By optimizing the  growth parameters,  we were able to obtain a maximum  superconducting transition temperature of   $ \sim $ 8 and 6  K for  films grown on MgO and SiO$ _{2} $, respectively. The surface morphology, crystal structure  and electrical characteristics of the films were investigated.  Films grown on MgO exhibited a  face centered cubic phase, while  an identical face centered cubic as well as a hexagonal phase coexisted in  films grown on oxidized silicon. Both these phases are known to be mononitrides.   We find that MgO is a better candidate for fabrication of high quality superconducting TaN thin films,  as it offers a higher $T_C  $ and lower resistivity, and is also more resistant to pressure fluctuation induced suppression of  $T_C $. 
\section*{Acknowledgments}
This research has been supported by Academy of Finland projects number 128532 and 260880. We thank Shen Boxuan for help with AFM measurements.  

%\bibliography{citations}

\begin{thebibliography}{35}%
\makeatletter
\providecommand \@ifxundefined [1]{%
 \@ifx{#1\undefined}
}%
\providecommand \@ifnum [1]{%
 \ifnum #1\expandafter \@firstoftwo
 \else \expandafter \@secondoftwo
 \fi
}%
\providecommand \@ifx [1]{%
 \ifx #1\expandafter \@firstoftwo
 \else \expandafter \@secondoftwo
 \fi
}%
\providecommand \natexlab [1]{#1}%
\providecommand \enquote  [1]{``#1''}%
\providecommand \bibnamefont  [1]{#1}%
\providecommand \bibfnamefont [1]{#1}%
\providecommand \citenamefont [1]{#1}%
\providecommand \href@noop [0]{\@secondoftwo}%
\providecommand \href [0]{\begingroup \@sanitize@url \@href}%
\providecommand \@href[1]{\@@startlink{#1}\@@href}%
\providecommand \@@href[1]{\endgroup#1\@@endlink}%
\providecommand \@sanitize@url [0]{\catcode `\\12\catcode `\$12\catcode
  `\&12\catcode `\#12\catcode `\^12\catcode `\_12\catcode `\%12\relax}%
\providecommand \@@startlink[1]{}%
\providecommand \@@endlink[0]{}%
\providecommand \url  [0]{\begingroup\@sanitize@url \@url }%
\providecommand \@url [1]{\endgroup\@href {#1}{\urlprefix }}%
\providecommand \urlprefix  [0]{URL }%
\providecommand \Eprint [0]{\href }%
\providecommand \doibase [0]{http://dx.doi.org/}%
\providecommand \selectlanguage [0]{\@gobble}%
\providecommand \bibinfo  [0]{\@secondoftwo}%
\providecommand \bibfield  [0]{\@secondoftwo}%
\providecommand \translation [1]{[#1]}%
\providecommand \BibitemOpen [0]{}%
\providecommand \bibitemStop [0]{}%
\providecommand \bibitemNoStop [0]{.\EOS\space}%
\providecommand \EOS [0]{\spacefactor3000\relax}%
\providecommand \BibitemShut  [1]{\csname bibitem#1\endcsname}%
\let\auto@bib@innerbib\@empty
%</preamble>
\bibitem [{\citenamefont {Poole}(2000)}]{Poole}%
  \BibitemOpen
  \bibfield  {author} {\bibinfo {author} {\bibfnamefont {C.~P.}\ \bibnamefont
  {Poole}},\ }\href@noop {} {\emph {\bibinfo {title} {Handbook of
  superconductivity}}}\ (\bibinfo  {publisher} {Academic Press},\ \bibinfo
  {address} {San Diego, CA},\ \bibinfo {year} {2000})\BibitemShut {NoStop}%
\bibitem [{\citenamefont {Wittmer}(1980)}]{wittmer:540}%
  \BibitemOpen
  \bibfield  {author} {\bibinfo {author} {\bibfnamefont {M.}~\bibnamefont
  {Wittmer}},\ }\href {\doibase 10.1063/1.91978} {\bibfield  {journal}
  {\bibinfo  {journal} {Applied Physics Letters}\ }\textbf {\bibinfo {volume}
  {37}},\ \bibinfo {pages} {540} (\bibinfo {year} {1980})}\BibitemShut
  {NoStop}%
\bibitem [{\citenamefont {Holloway}\ \emph {et~al.}(1992)\citenamefont
  {Holloway}, \citenamefont {Fryer}, \citenamefont {Cyril~Cabral},
  \citenamefont {Harper}, \citenamefont {Bailey},\ and\ \citenamefont
  {Kelleher}}]{holloway:5433}%
  \BibitemOpen
  \bibfield  {author} {\bibinfo {author} {\bibfnamefont {K.}~\bibnamefont
  {Holloway}}, \bibinfo {author} {\bibfnamefont {P.~M.}\ \bibnamefont {Fryer}},
  \bibinfo {author} {\bibfnamefont {J.}~\bibnamefont {Cyril~Cabral}}, \bibinfo
  {author} {\bibfnamefont {J.~M.~E.}\ \bibnamefont {Harper}}, \bibinfo {author}
  {\bibfnamefont {P.~J.}\ \bibnamefont {Bailey}}, \ and\ \bibinfo {author}
  {\bibfnamefont {K.~H.}\ \bibnamefont {Kelleher}},\ }\href {\doibase
  10.1063/1.350566} {\bibfield  {journal} {\bibinfo  {journal} {Journal of
  Applied Physics}\ }\textbf {\bibinfo {volume} {71}},\ \bibinfo {pages} {5433}
  (\bibinfo {year} {1992})}\BibitemShut {NoStop}%
\bibitem [{\citenamefont {Katz}\ \emph {et~al.}(1993)\citenamefont {Katz},
  \citenamefont {Pearton}, \citenamefont {Nakahara}, \citenamefont {Baiocchi},
  \citenamefont {Lane},\ and\ \citenamefont {Kovalchick}}]{katz:5208}%
  \BibitemOpen
  \bibfield  {author} {\bibinfo {author} {\bibfnamefont {A.}~\bibnamefont
  {Katz}}, \bibinfo {author} {\bibfnamefont {S.~J.}\ \bibnamefont {Pearton}},
  \bibinfo {author} {\bibfnamefont {S.}~\bibnamefont {Nakahara}}, \bibinfo
  {author} {\bibfnamefont {F.~A.}\ \bibnamefont {Baiocchi}}, \bibinfo {author}
  {\bibfnamefont {E.}~\bibnamefont {Lane}}, \ and\ \bibinfo {author}
  {\bibfnamefont {J.}~\bibnamefont {Kovalchick}},\ }\href {\doibase
  10.1063/1.353747} {\bibfield  {journal} {\bibinfo  {journal} {Journal of
  Applied Physics}\ }\textbf {\bibinfo {volume} {73}},\ \bibinfo {pages} {5208}
  (\bibinfo {year} {1993})}\BibitemShut {NoStop}%
\bibitem [{\citenamefont {Baba}\ and\ \citenamefont
  {Hatada}(1996)}]{Baba1996429}%
  \BibitemOpen
  \bibfield  {author} {\bibinfo {author} {\bibfnamefont {K.}~\bibnamefont
  {Baba}}\ and\ \bibinfo {author} {\bibfnamefont {R.}~\bibnamefont {Hatada}},\
  }\href {\doibase 10.1016/S0257-8972(95)02799-8} {\bibfield  {journal}
  {\bibinfo  {journal} {Surface and Coatings Technology}\ }\textbf {\bibinfo
  {volume} {84}},\ \bibinfo {pages} {429 } (\bibinfo {year}
  {1996})}\BibitemShut {NoStop}%
\bibitem [{\citenamefont {Engel}\ \emph {et~al.}(2012)\citenamefont {Engel},
  \citenamefont {Aeschbacher}, \citenamefont {Inderbitzin}, \citenamefont
  {Schilling}, \citenamefont {Il'in}, \citenamefont {Hofherr}, \citenamefont
  {Siegel}, \citenamefont {Semenov},\ and\ \citenamefont {H\"{u}bers}}]{SSPD}%
  \BibitemOpen
  \bibfield  {author} {\bibinfo {author} {\bibfnamefont {A.}~\bibnamefont
  {Engel}}, \bibinfo {author} {\bibfnamefont {A.}~\bibnamefont {Aeschbacher}},
  \bibinfo {author} {\bibfnamefont {K.}~\bibnamefont {Inderbitzin}}, \bibinfo
  {author} {\bibfnamefont {A.}~\bibnamefont {Schilling}}, \bibinfo {author}
  {\bibfnamefont {K.}~\bibnamefont {Il'in}}, \bibinfo {author} {\bibfnamefont
  {M.}~\bibnamefont {Hofherr}}, \bibinfo {author} {\bibfnamefont
  {M.}~\bibnamefont {Siegel}}, \bibinfo {author} {\bibfnamefont
  {A.}~\bibnamefont {Semenov}}, \ and\ \bibinfo {author} {\bibfnamefont
  {H.-W.}\ \bibnamefont {H\"{u}bers}},\ }\href {\doibase 10.1063/1.3684243}
  {\bibfield  {journal} {\bibinfo  {journal} {Applied Physics Letters}\
  }\textbf {\bibinfo {volume} {100}},\ \bibinfo {eid} {062601} (\bibinfo {year}
  {2012})}\BibitemShut {NoStop}%
\bibitem [{\citenamefont {Kaul}\ \emph {et~al.}(2001)\citenamefont {Kaul},
  \citenamefont {R.}, \citenamefont {Van~Duzer}, \citenamefont {Yu},
  \citenamefont {Newman},\ and\ \citenamefont {Rowell}}]{SNSAPL}%
  \BibitemOpen
  \bibfield  {author} {\bibinfo {author} {\bibfnamefont {A.~B.}\ \bibnamefont
  {Kaul}}, \bibinfo {author} {\bibfnamefont {W.~S.}\ \bibnamefont {R.}},
  \bibinfo {author} {\bibfnamefont {T.}~\bibnamefont {Van~Duzer}}, \bibinfo
  {author} {\bibfnamefont {L.}~\bibnamefont {Yu}}, \bibinfo {author}
  {\bibfnamefont {N.}~\bibnamefont {Newman}}, \ and\ \bibinfo {author}
  {\bibfnamefont {J.~M.}\ \bibnamefont {Rowell}},\ }\href@noop {} {\bibfield
  {journal} {\bibinfo  {journal} {Appl. Phys. Lett.}\ }\textbf {\bibinfo
  {volume} {78}},\ \bibinfo {pages} {99} (\bibinfo {year} {2001})}\BibitemShut
  {NoStop}%
\bibitem [{\citenamefont {Nevala}\ \emph {et~al.}(2009)\citenamefont {Nevala},
  \citenamefont {Maasilta}, \citenamefont {Senapati},\ and\ \citenamefont
  {Budhani}}]{minna}%
  \BibitemOpen
  \bibfield  {author} {\bibinfo {author} {\bibfnamefont {M.~R.}\ \bibnamefont
  {Nevala}}, \bibinfo {author} {\bibfnamefont {I.~J.}\ \bibnamefont
  {Maasilta}}, \bibinfo {author} {\bibfnamefont {K.}~\bibnamefont {Senapati}},
  \ and\ \bibinfo {author} {\bibfnamefont {R.~C.}\ \bibnamefont {Budhani}},\
  }\href@noop {} {\bibfield  {journal} {\bibinfo  {journal} {IEEE Trans. Appl.
  Supercond.}\ }\textbf {\bibinfo {volume} {19}},\ \bibinfo {pages} {253}
  (\bibinfo {year} {2009})}\BibitemShut {NoStop}%
\bibitem [{\citenamefont {Terao}(1971)}]{JJAP.10.248}%
  \BibitemOpen
  \bibfield  {author} {\bibinfo {author} {\bibfnamefont {N.}~\bibnamefont
  {Terao}},\ }\href {\doibase 10.1143/JJAP.10.248} {\bibfield  {journal}
  {\bibinfo  {journal} {Japanese Journal of Applied Physics}\ }\textbf
  {\bibinfo {volume} {10}},\ \bibinfo {pages} {248} (\bibinfo {year}
  {1971})}\BibitemShut {NoStop}%
\bibitem [{\citenamefont {Nie}\ \emph {et~al.}(2001)\citenamefont {Nie},
  \citenamefont {Xu}, \citenamefont {Wang}, \citenamefont {You}, \citenamefont
  {Yang}, \citenamefont {Ong}, \citenamefont {Li},\ and\ \citenamefont
  {Liew}}]{APA}%
  \BibitemOpen
  \bibfield  {author} {\bibinfo {author} {\bibfnamefont {H.}~\bibnamefont
  {Nie}}, \bibinfo {author} {\bibfnamefont {S.}~\bibnamefont {Xu}}, \bibinfo
  {author} {\bibfnamefont {S.}~\bibnamefont {Wang}}, \bibinfo {author}
  {\bibfnamefont {L.}~\bibnamefont {You}}, \bibinfo {author} {\bibfnamefont
  {Z.}~\bibnamefont {Yang}}, \bibinfo {author} {\bibfnamefont {C.}~\bibnamefont
  {Ong}}, \bibinfo {author} {\bibfnamefont {J.}~\bibnamefont {Li}}, \ and\
  \bibinfo {author} {\bibfnamefont {T.}~\bibnamefont {Liew}},\ }\href {\doibase
  10.1007/s003390000691} {\bibfield  {journal} {\bibinfo  {journal} {Applied
  Physics A}\ }\textbf {\bibinfo {volume} {73}},\ \bibinfo {pages} {229}
  (\bibinfo {year} {2001})}\BibitemShut {NoStop}%
\bibitem [{\citenamefont {Stampfl}\ and\ \citenamefont
  {Freeman}(2005)}]{PhysRevB.71.024111}%
  \BibitemOpen
  \bibfield  {author} {\bibinfo {author} {\bibfnamefont {C.}~\bibnamefont
  {Stampfl}}\ and\ \bibinfo {author} {\bibfnamefont {A.~J.}\ \bibnamefont
  {Freeman}},\ }\href {\doibase 10.1103/PhysRevB.71.024111} {\bibfield
  {journal} {\bibinfo  {journal} {Phys. Rev. B}\ }\textbf {\bibinfo {volume}
  {71}},\ \bibinfo {pages} {024111} (\bibinfo {year} {2005})}\BibitemShut
  {NoStop}%
\bibitem [{\citenamefont {Reichelt}\ \emph {et~al.}(1978)\citenamefont
  {Reichelt}, \citenamefont {Nellen},\ and\ \citenamefont
  {Mair}}]{reichelt:5284}%
  \BibitemOpen
  \bibfield  {author} {\bibinfo {author} {\bibfnamefont {K.}~\bibnamefont
  {Reichelt}}, \bibinfo {author} {\bibfnamefont {W.}~\bibnamefont {Nellen}}, \
  and\ \bibinfo {author} {\bibfnamefont {G.}~\bibnamefont {Mair}},\ }\href
  {\doibase 10.1063/1.324428} {\bibfield  {journal} {\bibinfo  {journal}
  {Journal of Applied Physics}\ }\textbf {\bibinfo {volume} {49}},\ \bibinfo
  {pages} {5284} (\bibinfo {year} {1978})}\BibitemShut {NoStop}%
\bibitem [{\citenamefont {Hardy}\ and\ \citenamefont
  {Hulm}(1954)}]{PhysRev.93.1004}%
  \BibitemOpen
  \bibfield  {author} {\bibinfo {author} {\bibfnamefont {G.~F.}\ \bibnamefont
  {Hardy}}\ and\ \bibinfo {author} {\bibfnamefont {J.~K.}\ \bibnamefont
  {Hulm}},\ }\href {\doibase 10.1103/PhysRev.93.1004} {\bibfield  {journal}
  {\bibinfo  {journal} {Phys. Rev.}\ }\textbf {\bibinfo {volume} {93}},\
  \bibinfo {pages} {1004} (\bibinfo {year} {1954})}\BibitemShut {NoStop}%
\bibitem [{\citenamefont {Kilbane}\ and\ \citenamefont
  {Habig}(1975)}]{kilbane:107}%
  \BibitemOpen
  \bibfield  {author} {\bibinfo {author} {\bibfnamefont {F.~M.}\ \bibnamefont
  {Kilbane}}\ and\ \bibinfo {author} {\bibfnamefont {P.~S.}\ \bibnamefont
  {Habig}},\ }\href {\doibase 10.1116/1.568734} {\bibfield  {journal} {\bibinfo
   {journal} {Journal of Vacuum Science and Technology}\ }\textbf {\bibinfo
  {volume} {12}},\ \bibinfo {pages} {107} (\bibinfo {year} {1975})}\BibitemShut
  {NoStop}%
\bibitem [{\citenamefont {{Boiko}}\ and\ \citenamefont
  {{Popova}}(1970)}]{Boiko}%
  \BibitemOpen
  \bibfield  {author} {\bibinfo {author} {\bibfnamefont {L.~G.}\ \bibnamefont
  {{Boiko}}}\ and\ \bibinfo {author} {\bibfnamefont {S.~V.}\ \bibnamefont
  {{Popova}}},\ }\href@noop {} {\bibfield  {journal} {\bibinfo  {journal}
  {Soviet Journal of Experimental and Theoretical Physics Letters}\ }\textbf
  {\bibinfo {volume} {12}},\ \bibinfo {pages} {70} (\bibinfo {year}
  {1970})}\BibitemShut {NoStop}%
\bibitem [{\citenamefont {Ensinger}\ \emph {et~al.}(1995)\citenamefont
  {Ensinger}, \citenamefont {Kiuchi},\ and\ \citenamefont
  {Satou}}]{ensinger:6630}%
  \BibitemOpen
  \bibfield  {author} {\bibinfo {author} {\bibfnamefont {W.}~\bibnamefont
  {Ensinger}}, \bibinfo {author} {\bibfnamefont {M.}~\bibnamefont {Kiuchi}}, \
  and\ \bibinfo {author} {\bibfnamefont {M.}~\bibnamefont {Satou}},\ }\href
  {\doibase 10.1063/1.359073} {\bibfield  {journal} {\bibinfo  {journal}
  {Journal of Applied Physics}\ }\textbf {\bibinfo {volume} {77}},\ \bibinfo
  {pages} {6630} (\bibinfo {year} {1995})}\BibitemShut {NoStop}%
\bibitem [{\citenamefont {Gerstenberg}\ and\ \citenamefont
  {Hall}(1964)}]{Gerstenberg01011964}%
  \BibitemOpen
  \bibfield  {author} {\bibinfo {author} {\bibfnamefont {D.}~\bibnamefont
  {Gerstenberg}}\ and\ \bibinfo {author} {\bibfnamefont {P.~M.}\ \bibnamefont
  {Hall}},\ }\href {\doibase 10.1149/1.2426296} {\bibfield  {journal} {\bibinfo
   {journal} {Journal of The Electrochemical Society}\ }\textbf {\bibinfo
  {volume} {111}},\ \bibinfo {pages} {936} (\bibinfo {year} {1964})},\  \BibitemShut {NoStop}%
\bibitem [{\citenamefont {Olowolafe}\ \emph {et~al.}(1992)\citenamefont
  {Olowolafe}, \citenamefont {Mogab}, \citenamefont {Gregory},\ and\
  \citenamefont {Kottke}}]{olowolafe:4099}%
  \BibitemOpen
  \bibfield  {author} {\bibinfo {author} {\bibfnamefont {J.~O.}\ \bibnamefont
  {Olowolafe}}, \bibinfo {author} {\bibfnamefont {C.~J.}\ \bibnamefont
  {Mogab}}, \bibinfo {author} {\bibfnamefont {R.~B.}\ \bibnamefont {Gregory}},
  \ and\ \bibinfo {author} {\bibfnamefont {M.}~\bibnamefont {Kottke}},\ }\href
  {\doibase 10.1063/1.352242} {\bibfield  {journal} {\bibinfo  {journal}
  {Journal of Applied Physics}\ }\textbf {\bibinfo {volume} {72}},\ \bibinfo
  {pages} {4099} (\bibinfo {year} {1992})}\BibitemShut {NoStop}%
\bibitem [{\citenamefont {Shin}\ \emph {et~al.}(2002)\citenamefont {Shin},
  \citenamefont {Kim}, \citenamefont {Gall}, \citenamefont {Greene},\ and\
  \citenamefont {Petrov}}]{Shin}%
  \BibitemOpen
  \bibfield  {author} {\bibinfo {author} {\bibfnamefont {C.-S.}\ \bibnamefont
  {Shin}}, \bibinfo {author} {\bibfnamefont {Y.-W.}\ \bibnamefont {Kim}},
  \bibinfo {author} {\bibfnamefont {D.}~\bibnamefont {Gall}}, \bibinfo {author}
  {\bibfnamefont {J.}~\bibnamefont {Greene}}, \ and\ \bibinfo {author}
  {\bibfnamefont {I.}~\bibnamefont {Petrov}},\ }\href {\doibase
  10.1016/S0040-6090(01)01618-2} {\bibfield  {journal} {\bibinfo  {journal}
  {Thin Solid Films}\ }\textbf {\bibinfo {volume} {402}},\ \bibinfo {pages}
  {172 } (\bibinfo {year} {2002})}\BibitemShut {NoStop}%
\bibitem [{\citenamefont {Rossnagel}(2002)}]{rossnagel:2328}%
  \BibitemOpen
  \bibfield  {author} {\bibinfo {author} {\bibfnamefont {S.~M.}\ \bibnamefont
  {Rossnagel}},\ }\href {\doibase 10.1116/1.1520556} {\bibfield  {journal}
  {\bibinfo  {journal} {J. Vac. Sci. Technol. B}\ }\textbf {\bibinfo {volume}
  {20}},\ \bibinfo {pages} {2328} (\bibinfo {year} {2002})}\BibitemShut
  {NoStop}%
\bibitem [{\citenamefont {Il$'$in}\ \emph {et~al.}(2012)\citenamefont
  {Il$'$in}, \citenamefont {Hofherr}, \citenamefont {Rall}, \citenamefont
  {Siegel}, \citenamefont {Semenov}, \citenamefont {Engel}, \citenamefont
  {Inderbitzin}, \citenamefont {Aeschbacher},\ and\ \citenamefont
  {Schilling}}]{Ilin}%
  \BibitemOpen
  \bibfield  {author} {\bibinfo {author} {\bibfnamefont {K.}~\bibnamefont
  {Il$'$in}}, \bibinfo {author} {\bibfnamefont {M.}~\bibnamefont {Hofherr}},
  \bibinfo {author} {\bibfnamefont {D.}~\bibnamefont {Rall}}, \bibinfo {author}
  {\bibfnamefont {M.}~\bibnamefont {Siegel}}, \bibinfo {author} {\bibfnamefont
  {A.}~\bibnamefont {Semenov}}, \bibinfo {author} {\bibfnamefont
  {A.}~\bibnamefont {Engel}}, \bibinfo {author} {\bibfnamefont
  {K.}~\bibnamefont {Inderbitzin}}, \bibinfo {author} {\bibfnamefont
  {A.}~\bibnamefont {Aeschbacher}}, \ and\ \bibinfo {author} {\bibfnamefont
  {A.}~\bibnamefont {Schilling}},\ }\href {\doibase 10.1007/s10909-011-0424-3}
  {\bibfield  {journal} {\bibinfo  {journal} {Journal of Low Temperature
  Physics}\ }\textbf {\bibinfo {volume} {167}},\ \bibinfo {pages} {809}
  (\bibinfo {year} {2012})}\BibitemShut {NoStop}%
\bibitem [{\citenamefont {Arranz}\ and\ \citenamefont
  {Palacio}(2005)}]{Arranz}%
  \BibitemOpen
  \bibfield  {author} {\bibinfo {author} {\bibfnamefont {A.}~\bibnamefont
  {Arranz}}\ and\ \bibinfo {author} {\bibfnamefont {C.}~\bibnamefont
  {Palacio}},\ }\href {\doibase 10.1007/s00339-004-3182-0} {\bibfield
  {journal} {\bibinfo  {journal} {Applied Physics A}\ }\textbf {\bibinfo
  {volume} {81}},\ \bibinfo {pages} {1405} (\bibinfo {year}
  {2005})}\BibitemShut {NoStop}%
\bibitem [{\citenamefont {Kawasaki}\ \emph {et~al.}()\citenamefont {Kawasaki},
  \citenamefont {Namba},\ and\ \citenamefont {Suda}}]{HK}%
  \BibitemOpen
  \bibfield  {author} {\bibinfo {author} {\bibfnamefont {H.}~\bibnamefont
  {Kawasaki}}, \bibinfo {author} {\bibfnamefont {K.~D.~J.}\ \bibnamefont
  {Namba}}, \ and\ \bibinfo {author} {\bibfnamefont {Y.}~\bibnamefont {Suda}},\
  }\href {\doibase 10.1557/PROC-617-J3.22} {\bibfield  {journal} {\bibinfo
  {journal} {MRS Proceedings}\ }\textbf {\bibinfo {volume} {617}},\
  10.1557/PROC-617-J3.22}\BibitemShut {NoStop}%
\bibitem [{\citenamefont {Matenoglou}\ \emph {et~al.}(2008)\citenamefont
  {Matenoglou}, \citenamefont {Koutsokeras}, \citenamefont {Lekka},
  \citenamefont {Abadias}, \citenamefont {Camelio}, \citenamefont
  {Evangelakis}, \citenamefont {Kosmidis},\ and\ \citenamefont
  {Patsalas}}]{matenoglou}%
  \BibitemOpen
  \bibfield  {author} {\bibinfo {author} {\bibfnamefont {G.~M.}\ \bibnamefont
  {Matenoglou}}, \bibinfo {author} {\bibfnamefont {L.~E.}\ \bibnamefont
  {Koutsokeras}}, \bibinfo {author} {\bibfnamefont {C.~E.}\ \bibnamefont
  {Lekka}}, \bibinfo {author} {\bibfnamefont {G.}~\bibnamefont {Abadias}},
  \bibinfo {author} {\bibfnamefont {S.}~\bibnamefont {Camelio}}, \bibinfo
  {author} {\bibfnamefont {G.~A.}\ \bibnamefont {Evangelakis}}, \bibinfo
  {author} {\bibfnamefont {C.}~\bibnamefont {Kosmidis}}, \ and\ \bibinfo
  {author} {\bibfnamefont {P.}~\bibnamefont {Patsalas}},\ }\href {\doibase
  10.1063/1.3043882} {\bibfield  {journal} {\bibinfo  {journal} {Journal of
  Applied Physics}\ }\textbf {\bibinfo {volume} {104}},\ \bibinfo {eid}
  {124907} (\bibinfo {year} {2008})}\BibitemShut {NoStop}%
\bibitem [{\citenamefont {Chaudhuri}\ \emph {et~al.}(2011)\citenamefont
  {Chaudhuri}, \citenamefont {Nevala}, \citenamefont {Hakkarainen},
  \citenamefont {Niemi},\ and\ \citenamefont {Maasilta}}]{nbn}%
  \BibitemOpen
  \bibfield  {author} {\bibinfo {author} {\bibfnamefont {S.}~\bibnamefont
  {Chaudhuri}}, \bibinfo {author} {\bibfnamefont {M.}~\bibnamefont {Nevala}},
  \bibinfo {author} {\bibfnamefont {T.}~\bibnamefont {Hakkarainen}}, \bibinfo
  {author} {\bibfnamefont {T.}~\bibnamefont {Niemi}}, \ and\ \bibinfo {author}
  {\bibfnamefont {I.}~\bibnamefont {Maasilta}},\ }\href {\doibase
  10.1109/TASC.2010.2081656} {\bibfield  {journal} {\bibinfo  {journal}
  {IEEE Trans. Appl. Supercond.}\ }\textbf {\bibinfo
  {volume} {21}},\ \bibinfo {pages} {143 } (\bibinfo {year}
  {2011})}\BibitemShut {NoStop}%
\bibitem [{\citenamefont {Chaudhuri}\ \emph {et~al.}()\citenamefont
  {Chaudhuri}, \citenamefont {Nevala},\ and\ \citenamefont {Maasilta}}]{NbN2}%
  \BibitemOpen
  \bibfield  {author} {\bibinfo {author} {\bibfnamefont {S.}~\bibnamefont
  {Chaudhuri}}, \bibinfo {author} {\bibfnamefont {M.~R.}\ \bibnamefont
  {Nevala}}, \ and\ \bibinfo {author} {\bibfnamefont {I.~J.}\ \bibnamefont
  {Maasilta}},\ }\href@noop {} {\bibinfo  {journal} {Submitted, arXiv:1301.6003}\ }\BibitemShut
  {NoStop}%
\bibitem [{\citenamefont {Nevala}\ \emph {et~al.}(2012)\citenamefont {Nevala},
  \citenamefont {Chaudhuri}, \citenamefont {Halkosaari}, \citenamefont
  {Karvonen},\ and\ \citenamefont {Maasilta}}]{Nb}%
  \BibitemOpen
\bibfield  {journal} {  }\bibfield  {author} {\bibinfo {author} {\bibfnamefont
  {M.~R.}\ \bibnamefont {Nevala}}, \bibinfo {author} {\bibfnamefont
  {S.}~\bibnamefont {Chaudhuri}}, \bibinfo {author} {\bibfnamefont
  {J.}~\bibnamefont {Halkosaari}}, \bibinfo {author} {\bibfnamefont {J.~T.}\
  \bibnamefont {Karvonen}}, \ and\ \bibinfo {author} {\bibfnamefont {I.~J.}\
  \bibnamefont {Maasilta}},\ }\href {\doibase 10.1063/1.4751355} {\bibfield
  {journal} {\bibinfo  {journal} {Applied Physics Letters}\ }\textbf {\bibinfo
  {volume} {101}},\ \bibinfo {eid} {112601} (\bibinfo {year}
  {2012})}\BibitemShut {NoStop}%
\bibitem [{\citenamefont {Giazotto}\ \emph {et~al.}(2006)\citenamefont
  {Giazotto}, \citenamefont {Heikkil\"a}, \citenamefont {Luukanen},
  \citenamefont {Savin},\ and\ \citenamefont {Pekola}}]{RMP}%
  \BibitemOpen
  \bibfield  {author} {\bibinfo {author} {\bibfnamefont {F.}~\bibnamefont
  {Giazotto}}, \bibinfo {author} {\bibfnamefont {T.~T.}\ \bibnamefont
  {Heikkil\"a}}, \bibinfo {author} {\bibfnamefont {A.}~\bibnamefont
  {Luukanen}}, \bibinfo {author} {\bibfnamefont {A.~M.}\ \bibnamefont {Savin}},
  \ and\ \bibinfo {author} {\bibfnamefont {J.~P.}\ \bibnamefont {Pekola}},\
  }\href {\doibase 10.1103/RevModPhys.78.217} {\bibfield  {journal} {\bibinfo
  {journal} {Rev. Mod. Phys.}\ }\textbf {\bibinfo {volume} {78}},\ \bibinfo
  {pages} {217} (\bibinfo {year} {2006})}\BibitemShut {NoStop}%
\bibitem [{\citenamefont {Muhonen}\ \emph {et~al.}(2012)\citenamefont
  {Muhonen}, \citenamefont {Meschke},\ and\ \citenamefont {Pekola}}]{muhonen}%
  \BibitemOpen
  \bibfield  {author} {\bibinfo {author} {\bibfnamefont {J.~T.}\ \bibnamefont
  {Muhonen}}, \bibinfo {author} {\bibfnamefont {M.}~\bibnamefont {Meschke}}, \
  and\ \bibinfo {author} {\bibfnamefont {J.~P.}\ \bibnamefont {Pekola}},\
  }\href@noop {} {\bibfield  {journal} {\bibinfo  {journal} {Rep. Prog. Phys.}\
  }\textbf {\bibinfo {volume} {75}},\ \bibinfo {pages} {046501} (\bibinfo
  {year} {2012})}\BibitemShut {NoStop}%
\bibitem [{\citenamefont {Li}\ \emph {et~al.}(2006)\citenamefont {Li},
  \citenamefont {Er-Wu}, \citenamefont {Guo-Hua}, \citenamefont {Wen-Ran},
  \citenamefont {Wei-Chao}, \citenamefont {Guang-Liang}, \citenamefont
  {Gu-Ling}, \citenamefont {Song-Hua}, \citenamefont {Chi-Zi},\ and\
  \citenamefont {Si-Ze}}]{Li}%
  \BibitemOpen
  \bibfield  {author} {\bibinfo {author} {\bibfnamefont {L.}~\bibnamefont
  {Li}}, \bibinfo {author} {\bibfnamefont {N.}~\bibnamefont {Er-Wu}}, \bibinfo
  {author} {\bibfnamefont {L.}~\bibnamefont {Guo-Hua}}, \bibinfo {author}
  {\bibfnamefont {F.}~\bibnamefont {Wen-Ran}}, \bibinfo {author} {\bibfnamefont
  {G.}~\bibnamefont {Wei-Chao}}, \bibinfo {author} {\bibfnamefont
  {C.}~\bibnamefont {Guang-Liang}}, \bibinfo {author} {\bibfnamefont
  {Z.}~\bibnamefont {Gu-Ling}}, \bibinfo {author} {\bibfnamefont
  {F.}~\bibnamefont {Song-Hua}}, \bibinfo {author} {\bibfnamefont
  {L.}~\bibnamefont {Chi-Zi}}, \ and\ \bibinfo {author} {\bibfnamefont
  {Y.}~\bibnamefont {Si-Ze}},\ }\href
  {http://cpl.iphy.ac.cn/EN/abstract/article_39603.shtml} {\bibfield  {journal}
  {\bibinfo  {journal} {Chinese Physics Letters}\ }\textbf {\bibinfo {volume}
  {23}},\ \bibinfo {eid} {3018} (\bibinfo {year} {2006})}\BibitemShut {NoStop}%
\bibitem [{\citenamefont {Breznay}\ \emph {et~al.}(2012)\citenamefont
  {Breznay}, \citenamefont {Michaeli}, \citenamefont {Tikhonov}, \citenamefont
  {Finkel'stein}, \citenamefont {Tendulkar},\ and\ \citenamefont
  {Kapitulnik}}]{PhysRevB.86.014514}%
  \BibitemOpen
  \bibfield  {author} {\bibinfo {author} {\bibfnamefont {N.~P.}\ \bibnamefont
  {Breznay}}, \bibinfo {author} {\bibfnamefont {K.}~\bibnamefont {Michaeli}},
  \bibinfo {author} {\bibfnamefont {K.~S.}\ \bibnamefont {Tikhonov}}, \bibinfo
  {author} {\bibfnamefont {A.~M.}\ \bibnamefont {Finkel'stein}}, \bibinfo
  {author} {\bibfnamefont {M.}~\bibnamefont {Tendulkar}}, \ and\ \bibinfo
  {author} {\bibfnamefont {A.}~\bibnamefont {Kapitulnik}},\ }\href {\doibase
  10.1103/PhysRevB.86.014514} {\bibfield  {journal} {\bibinfo  {journal} {Phys.
  Rev. B}\ }\textbf {\bibinfo {volume} {86}},\ \bibinfo {pages} {014514}
  (\bibinfo {year} {2012})}\BibitemShut {NoStop}%
\bibitem [{\citenamefont {Kang}\ \emph {et~al.}(2011)\citenamefont {Kang},
  \citenamefont {Jin}, \citenamefont {Liu}, \citenamefont {Jia}, \citenamefont
  {Chen}, \citenamefont {Ji}, \citenamefont {Xu}, \citenamefont {Wu},
  \citenamefont {Mi}, \citenamefont {Pimenov}, \citenamefont {Wu},\ and\
  \citenamefont {Wang}}]{kang:033908}%
  \BibitemOpen
  \bibfield  {author} {\bibinfo {author} {\bibfnamefont {L.}~\bibnamefont
  {Kang}}, \bibinfo {author} {\bibfnamefont {B.~B.}\ \bibnamefont {Jin}},
  \bibinfo {author} {\bibfnamefont {X.~Y.}\ \bibnamefont {Liu}}, \bibinfo
  {author} {\bibfnamefont {X.~Q.}\ \bibnamefont {Jia}}, \bibinfo {author}
  {\bibfnamefont {J.}~\bibnamefont {Chen}}, \bibinfo {author} {\bibfnamefont
  {Z.~M.}\ \bibnamefont {Ji}}, \bibinfo {author} {\bibfnamefont {W.~W.}\
  \bibnamefont {Xu}}, \bibinfo {author} {\bibfnamefont {P.~H.}\ \bibnamefont
  {Wu}}, \bibinfo {author} {\bibfnamefont {S.~B.}\ \bibnamefont {Mi}}, \bibinfo
  {author} {\bibfnamefont {A.}~\bibnamefont {Pimenov}}, \bibinfo {author}
  {\bibfnamefont {Y.~J.}\ \bibnamefont {Wu}}, \ and\ \bibinfo {author}
  {\bibfnamefont {B.~G.}\ \bibnamefont {Wang}},\ }\href {\doibase
  10.1063/1.3518037} {\bibfield  {journal} {\bibinfo  {journal} {Journal of
  Applied Physics}\ }\textbf {\bibinfo {volume} {109}},\ \bibinfo {eid}
  {033908} (\bibinfo {year} {2011})}\BibitemShut {NoStop}%
\bibitem [{\citenamefont {Matsuo}\ \emph {et~al.}(1974)\citenamefont {Matsuo},
  \citenamefont {Sugiura},\ and\ \citenamefont {Noguchi}}]{T1s}%
  \BibitemOpen
  \bibfield  {author} {\bibinfo {author} {\bibfnamefont {S.}~\bibnamefont
  {Matsuo}}, \bibinfo {author} {\bibfnamefont {H.}~\bibnamefont {Sugiura}}, \
  and\ \bibinfo {author} {\bibfnamefont {S.}~\bibnamefont {Noguchi}},\ }\href
  {\doibase 10.1007/BF00654622} {\bibfield  {journal} {\bibinfo  {journal}
  {Journal of Low Temperature Physics}\ }\textbf {\bibinfo {volume} {15}},\
  \bibinfo {pages} {481} (\bibinfo {year} {1974})}\BibitemShut {NoStop}%
\bibitem [{\citenamefont {Ohshima}\ \emph {et~al.}(1976)\citenamefont
  {Ohshima}, \citenamefont {Kuroishi},\ and\ \citenamefont
  {Fujita}}]{JPSJ.41.1234}%
  \BibitemOpen
  \bibfield  {author} {\bibinfo {author} {\bibfnamefont {K.}~\bibnamefont
  {Ohshima}}, \bibinfo {author} {\bibfnamefont {T.}~\bibnamefont {Kuroishi}}, \
  and\ \bibinfo {author} {\bibfnamefont {T.}~\bibnamefont {Fujita}},\ }\href
  {\doibase 10.1143/JPSJ.41.1234} {\bibfield  {journal} {\bibinfo  {journal}
  {Journal of the Physical Society of Japan}\ }\textbf {\bibinfo {volume}
  {41}},\ \bibinfo {pages} {1234} (\bibinfo {year} {1976})}\BibitemShut
  {NoStop}%
\bibitem [{\citenamefont {Abeles}\ \emph {et~al.}(1966)\citenamefont {Abeles},
  \citenamefont {Cohen},\ and\ \citenamefont {Cullen}}]{PhysRevLett.17.632}%
  \BibitemOpen
  \bibfield  {author} {\bibinfo {author} {\bibfnamefont {B.}~\bibnamefont
  {Abeles}}, \bibinfo {author} {\bibfnamefont {R.~W.}\ \bibnamefont {Cohen}}, \
  and\ \bibinfo {author} {\bibfnamefont {G.~W.}\ \bibnamefont {Cullen}},\
  }\href {\doibase 10.1103/PhysRevLett.17.632} {\bibfield  {journal} {\bibinfo
  {journal} {Phys. Rev. Lett.}\ }\textbf {\bibinfo {volume} {17}},\ \bibinfo
  {pages} {632} (\bibinfo {year} {1966})}\BibitemShut {NoStop}%
\end{thebibliography}

%merlin.mbs apsrev4-1.bst 2010-07-25 4.21a (PWD, AO, DPC) hacked
%Control: key (0)
%Control: author (8) initials jnrlst
%Control: editor formatted (1) identically to author
%Control: production of article title (-1) disabled
%Control: page (0) single
%Control: year (1) truncated
%Control: production of eprint (0) enabled
%

\end{document}